\documentstyle[prl,aps,epsf]{revtex} 
\twocolumn 

\newcommand{\etal}{{\em et al.{}}}
\newcommand{\be}{\begin{equation}}
\newcommand{\ee}{\end{equation}}
\newcommand{\bea}{\begin{eqnarray}}
\newcommand{\eea}{\end{eqnarray}}
\newcommand{\de}{\mbox{d}}     
\newcommand{\PostScript}[6]{
\begin{figure}[htb]
\vspace{#2cm}
\begin{center}
\epsfysize=#1cm
\leavevmode
\epsfbox{#3}
\par
\end{center}
\mycaption{#4}{figure}{#5}{#6}
\vspace{-0.4cm}
\end{figure}
}
\newcommand{\mycaption}[4]
              {\begin{center}
               \vspace{#1cm}
               \parbox{8.6cm}{
                \caption[#2]{\renewcommand{\baselinestretch}{1} \small \em #3}
                \label{#4}
               }
               \end{center}
              }

\begin{document}
\draft

\title{Classical orbit bifurcation and quantum interference
       in mesoscopic magnetoconductance}
\author{J. Blaschke and M. Brack}
\address{Institut f\"ur Theoretische Physik, Universit\"at Regensburg,
         D-93040 Regensburg, Germany}
\date{\today}
\maketitle

\widetext
\vspace*{-5.0cm}
\begin{flushright}
 {\bf TPR-99-07}
\end{flushright}
\vspace*{3.5cm}
\narrowtext

\begin{abstract}
We study the magnetoconductance of electrons through a mesoscopic
channel with antidots. Through quantum interference effects, the
conductance maxima as functions of the magnetic field strength and
the antidot radius (regulated by the applied gate voltage) exhibit
characteristic dislocations that have been observed experimentally.
Using the semiclassical periodic orbit theory, we relate these
dislocations directly to bifurcations of the leading classes of
periodic orbits.
\end{abstract}


Since it has become feasible to laterally confine a two-dimensional
electron gas (2DEG) on length scales considerably smaller than
the mean-free path of the electrons, the connection between classical 
and quantum mechanics has gained increasing renewed interest.
Therefore, many experimental and theoretical investigations have
recently been focused on the onset of quantum interference effects in
mesoscopic ballistic devices. A well-adapted theoretical tool for
this regime is the semiclassical approach that approximates quantum
mechanics to leading orders in $\hbar$. It is conceptually quite
remarkable because it links quantum interference effects to purely
classical phase-space dynamics. The so-called trace formula,
originally developed by Gutzwiller \cite{Gutzwiller:All} for the
density of states of a system with only isolated orbits in phase space,
has been extended to systems with continuous
symmetries \cite{Strut:Mag,Creagh:Cont} (see Ref.~\cite{Brack:Buch}
for further literature) and for other physical properties such as
conductance \cite{Richter:EPL29,Hackenbroich:Gxx} or magnetic
susceptibility \cite{RUJ,Tanaka:AP268}.
Many quantum interference effects observed in mesoscopic systems could
successfully be explained by the interference of a few classical
periodic orbits, such as the Shubnikov-de-Haas oscillations of the free
2DEG \cite{Richter:EPL29,Hackenbroich:Gxx}, the magnetoconductance
oscillations of a 2DEG in an antidot superlattice
\cite{Richter:EPL29,Hackenbroich:Gxx,Weiss:PRL70,Lorke:PRB44} or a
large circular quantum dot \cite{Reimann:ZPB101}, or the current
oscillations in a resonant tunneling diode (RTD)
\cite{RTD:all}. (See Ref.\ \cite{Brack:Buch} for examples from nuclear
and metal cluster physics.)

Real physical systems are usually neither integrable nor fully chaotic,
but exhibit mixed phase-space dynamics. Upon variation of an external 
parameter (e.g., deformation, magnetic field strength, or energy), 
bifurcations of periodic orbits typically occur, whereby new orbits
are born and/or old orbits vanish. In the RTD \cite{RTD:all}, 
period-doubling bifurcations were found to be responsible for a period
doubling in the oscillations of the observed I-V curves. In 
superdeformed nuclei \cite{Matsuy} and in the elliptic billiard 
\cite{Mag:bif}, period-doubling bifurcations dominate the 
quantum shell structure locally through new-born orbit 
families whose amplitudes are of relative order $1/\hbar$. 
Here we discuss a different
mechanism through which orbit bifurcations manifest themselves in the
magnetoconductance of a mesoscopic device, a narrow channel with
central antidots. We present a semiclassical interpretation of
dislocations in the conductance maxima as functions 
of antidot diameter and magnetic field strength and relate them 
to bifurcations of the leading classes of periodic orbits. We will
show that their effect is neither due to their leading order  
in $\hbar$ nor to period doubling, but to a subtle interference
of different orbit generations with comparable periods.

For the semiclassical description of the conductance we follow the
approach of Refs.\ \cite{Richter:EPL29,Hackenbroich:Gxx}. The smooth
part of the conductance $G_{xx}$ (in the direction $x$ of the electric
current) can be described by the classical Kubo formula, whereas its 
oscillating part $\delta G_{xx}$ is approximated in terms of periodic 
orbits (po):
 \bea
  \label{traceFormula}
  && \delta G_{xx}  = \nonumber \\
  && \frac{1}{\ell^2} \frac{4e^2}{h} \sum_{\rm po}{\cal C}_{xx}
     \frac{R_{\rm po}(\tau_\beta) \, F_{\rm po}(\tau_s)}
      {|{\rm Det}(\widetilde{\rm M}_{po}-{\bf 1})|^{1/2}}
     \cos\left(\frac{S_{po}}{\hbar}-\mu_{po}\frac{\pi}{2}\right).
 \eea
Here $S_{po}$ is the action (evaluated at the Fermi energy $E_F$),
$\mu_{po}$ the Maslov index, and ${\widetilde {\rm M}}_{po}$ the
stability matrix of each periodic orbit \cite{Gutzwiller:All}. The
temperature $T$ is included in the factor $R_{\rm po}(\tau_\beta) =
(T_{po}/\tau_\beta)/\sinh(T_{po}/\tau_\beta)$ involving the 
(primitive) time period $T_{po}$ and the scattering time 
$\tau_\beta=\hbar/(\pi kT)\approx 2.4*10^{-11}{\rm s}$. Damping 
due to a finite mean free path is given by $ F_{\rm po}(\tau_s) = 
e^{-T_{po}/(2\tau_s)}$, where $\tau_s=m^\star\mu/e \approx 
3.8*10^{-11}{\rm s}$ is the scattering time extracted 
from the experimental mobility $\mu$. $\ell\simeq 1\mu$m is the 
characteristic length of the active region, and ${\cal C}_{xx}$ 
is the velocity-velocity correlation function of the periodic 
orbit, defined by
 \be
  {\cal C}_{xx}=\int_0^\infty \de t \, e^{-t/\tau_s}
                \int_0^{T_{po}}\de \tau \, v_x(\tau) \, v_x(t+\tau)\,.
 \ee

Eq.\ (\ref{traceFormula}), as well as the standard trace formulae for the
density of states, diverges at bifurcation points where two (or more) 
stationary points of the action coalesce and the stationary-phase 
approximation \cite{Gutzwiller:All} in the trace integral leads to 
${\rm Det} ({\widetilde{\rm M}}_{po}-{\bf 1})=0$. This can be locally 
overcome by expanding the action into higher-order normal forms 
\cite{Almeida:JPA20}. The simultaneous requirement of asymptotically
reaching the Gutzwiller amplitudes far from the bifurcation points leads 
to uniform approximations which were developed systematically by 
Sieber and Schomerus \cite{Sieber:Bifs}. At the critical points 
the amplitudes are increased by a factor $\hbar^{-\delta}$, where
$\delta$ depends on the type of the bifurcation\cite{foot:delta}. In
the classical limit $\hbar/S \rightarrow 0$, bifurcations therefore 
may dominate the quantum oscillations. In real systems
$\hbar/S$ is, however, finite so that the other pre\-factors in the trace
formula may compensate the factor $(S/\hbar)^{\delta}$ to a degree
that depends on the specific system. 
Note that the uniform approximations of Refs.\ 
\cite{Sieber:Bifs} are restricted to isolated bifurcations; a general 
treatment of bifurcations of higher codimension (i.e., bifurcations
of bifurcations) is still lacking \cite{Schom:codim2}. 
We employ a slightly modified version of the uniform treatment of Ref.\
\cite{Sieber:Bifs}, incorporating the discrete symmetries
of the present system. To include effects of 
complex `ghost orbits' not available in our calculations, we use the 
local approximation of their contribution derived
from the numerical information at the bifurcation points.
(For the technical details we refer to a forthcoming
extended publication.) 

The device investigated here consists of electrostatic gates confining
a high-mobility 2DEG in a GaAs/GaAlAs heterostructure (see Fig.\
\ref{Device}). The 2DEG was 82nm beneath the surface, its electron
density was $\approx 3.47*10^{15} {\rm m}^{-2}$, and the mobility about $100
{\rm m}^2 V^{-1}{\rm s}^{-1}$. Four metallic gates are used to define a long,
narrow channel ($5\mu$m$\times 1\mu$m). These and two circular
antidot gates are contacted individually. Details about the device are
presented in \cite{Gould:PRB51,Gould:CJP74,Kirczenow:PRB56} and the
references cited therein. All measurements were taken at $T \approx 100$
mK using standard low-excitation AC techniques.

\PostScript{3.2}{-0.4}{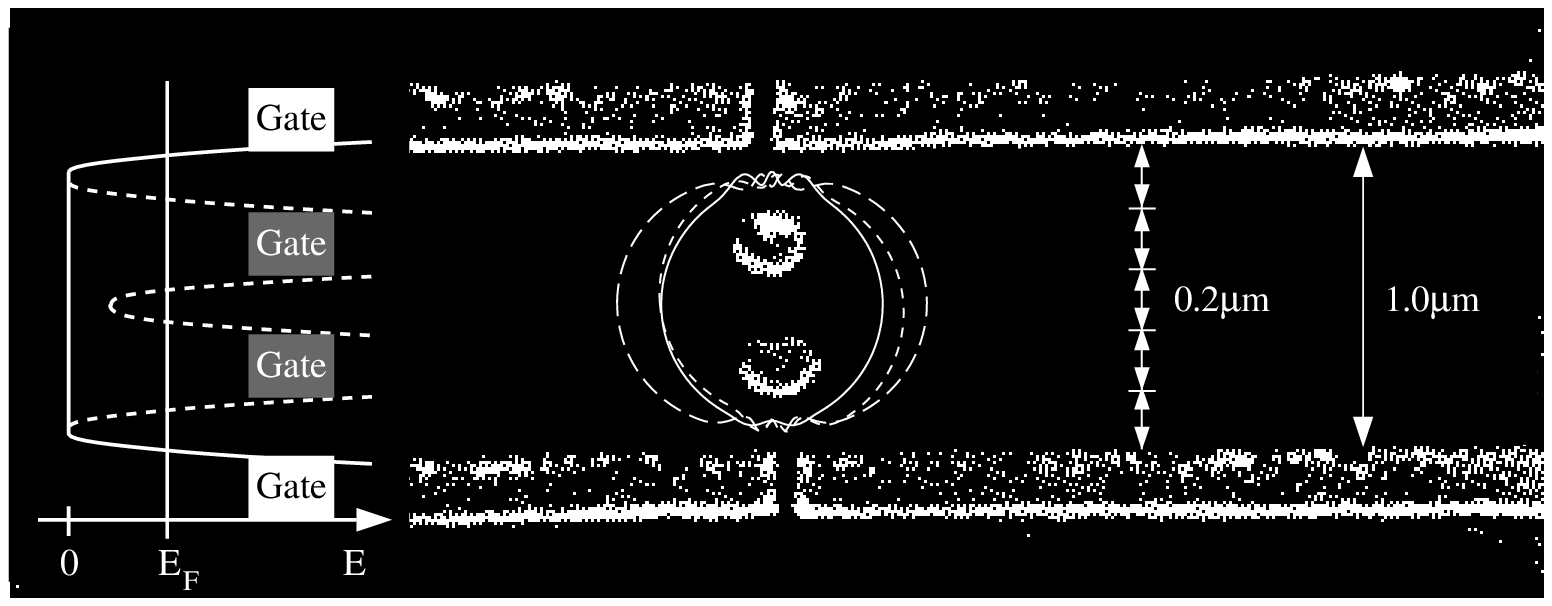}{-0.1}
           {SEM photograph of the gate structure \cite{Gould:PRB51} 
           (without contacts). 
           Left: Model potential used for the calculations. 
           Center: Typical periodic orbits. Note that some of them 
           break the discrete symmetries of the potential.
           }
           {Device}

The dots in Fig.\ \ref{Results}(a) show the experimental maximum 
positions of $\delta G_{xx}$ as functions of magnetic
field $B$ and antidot gate voltage $V_g$ \cite{Kirczenow:PRB56}. 
The nearly equally spaced maxima and their
shift to higher $B$ for decreasing antidot diameter can be understood
in analogy to the Aharonov-Bohm (AB) effect, if the AB ring is
identified with cyclotron orbits around the antidots. Extracting the
effective area from the experimental data yields a diameter between
$0.76\mu$m and $0.86\mu$m, which is consistent with the device
dimensions. The dislocations of the peak positions (see the boxes in
Fig.~\ref{Results}), however, cannot be understood within this simple
picture. They have been qualitatively reproduced in a quantum
calculation by Kirczenov \etal~\cite{Kirczenow:PRB56}.

\PostScript{5.7}{-0.7}{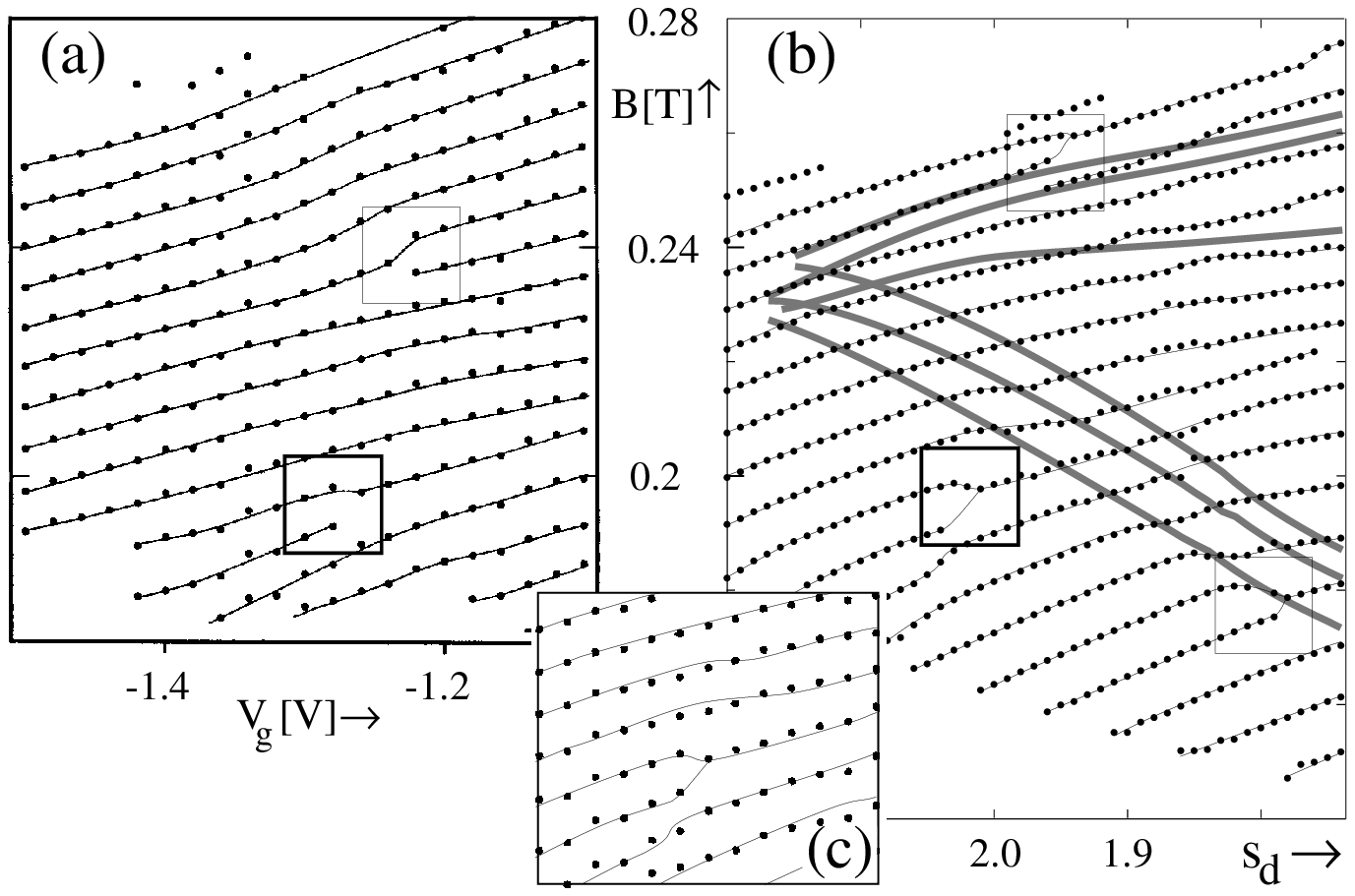}{-0.05}
            {Maximum positions of $\delta G_{xx}$ versus magnetic field
            strength and gate voltage (antidot size). (a) Dots from
            experiment, connected with lines to guide the eye 
            (reproduced with kind permission of the authors
            \cite{Kirczenow:PRB56}). (b) Semiclassical
            results. The gray-shaded lines correspond to the loci
            of orbit bifurcations. (c) Behavior near a dislocation
            (see text). Dots: experiment; lines: semiclassical results.
            }
            {Results}

Our objective is to decide if these dislocations and the variation of
the spacings between the maxima can be understood semiclassically
(which was doubted in Refs.\ \cite{Gould:PRB51,Kirczenow:PRB56}). 
For the effective one-electron model potential we follow 
essentially Kirczenov \etal\ \cite{Kirczenow:PRB56} who assumed a
parabolic shape $V(r)= E_F \left[r/a_0-(1+s)\right]^2$ for $r<a_0(1+s)$
and $V(r)= 0$ otherwise. Here $r$ denotes the distance to the gate, and
$a_0$ is the length scale over which the potential falls from $E_F$
to 0, i.e., the diffuseness of the potential. $s$ is a
dimensionless parameter modeling the depletion width around the gates.
We use $a_0=0.05 \mu m$ and $s=s_c=1$ for the gates defining the
channel throughout this paper. The depletion width $s_d$ of the antidot
gates is varied between 1.5 and 2.2. According to Ref.\
\cite{Kirczenow:PRB56}, this corresponds to an effective
antidot diameter of $\approx 0.35 \mu$m to $0.42 \mu$m.

Following Eckhardt and Wintgen \cite{Eckhardt:JPA91}, we
numerically integrate simultaneously the classical equations of
motion and the reduced (2D) stability matrix $\widetilde {\rm M}$. 
The orbits are converged to periodicity using the information 
provided by $\widetilde {\rm M}$. They are followed through varying 
$B$ fields and antidot diameters using an adaptive extrapolation 
scheme. We find a large variety of distinct periodic orbits, many of 
them breaking the symmetry of the potential. 
Some typical examples are shown in Figs.\ \ref{Device} and 
\ref{Bifurcation}. We have included over 60 orbits (not counting 
their symmetry-related partners). Their actions,
velocity-velocity correlation functions and periods were evaluated
numerically. The Maslov index was determined similarly to
Ref.\ \cite{Eckhardt:JPA91}.

Since the orbit bifurcations are of leading order in $\hbar$, we now
want to check if they have an increased influence on the amplitude of
the conductance oscillations.
In Fig.\ \ref{Bifurcation}(a) we show the quantity Tr$\widetilde{\rm M}$ 
of four typical periodic orbits (shown to the right) taking part in two
successive bifurcations (where Tr$\widetilde{\rm M}=2$) under variation
of the magnetic field strength $B$. The left one is a tangent
bifurcation, the right one a pitchfork bifurcation.

\PostScript{5.3}{-0.4}{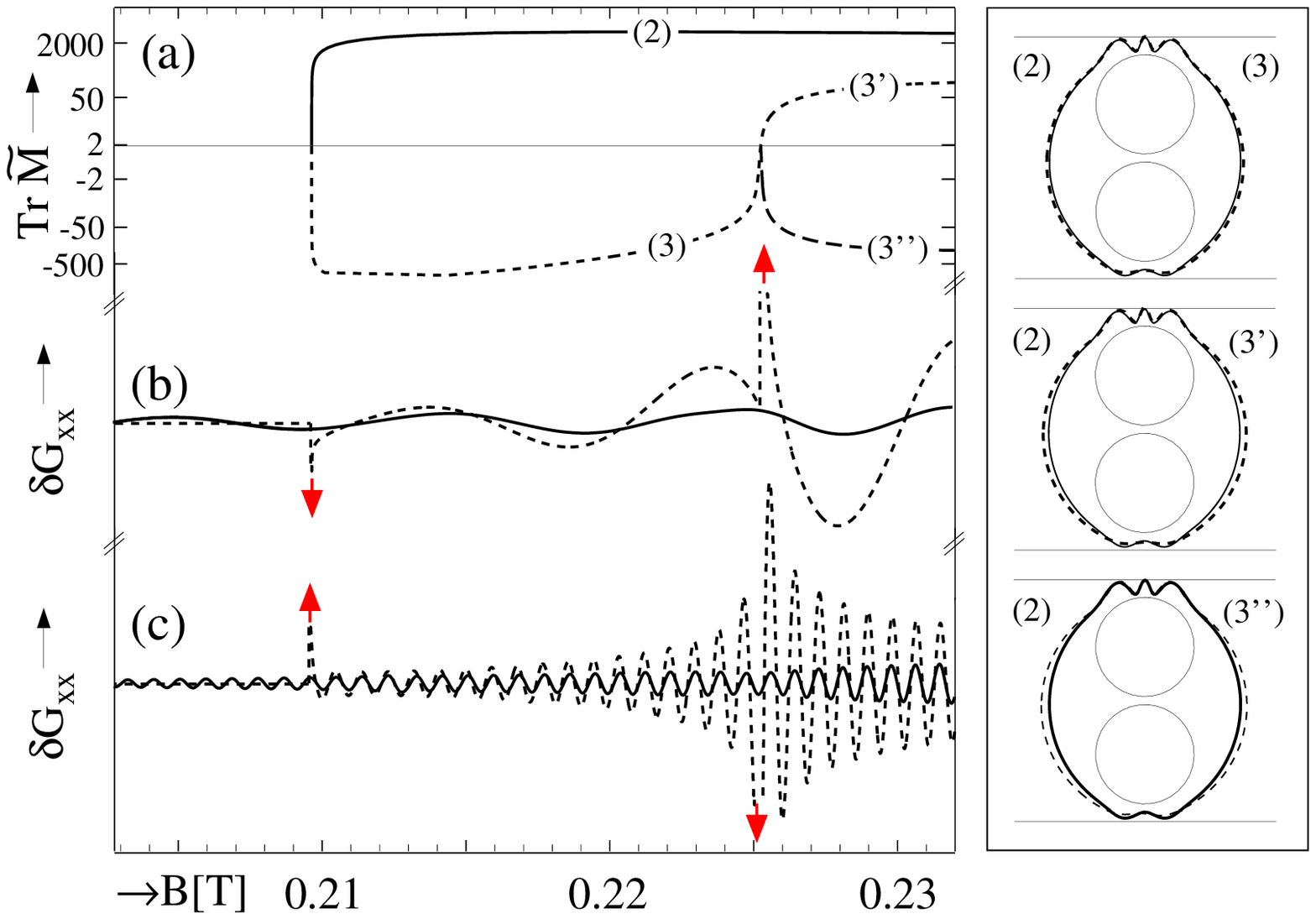}{-0.2}
            {Right: Four typical orbits (2: solid line; 3, 3', and 3'':
            dashed lines) involved in two successive bifurcations.
            {\it Left:} (a) Trace ${\rm Tr}\widetilde{\rm M}$ (note the 
            nonlinear scale) versus magnetic field $B$.
            (b) Contribution of all orbits to $\delta G_{xx}$. 
            Dotted line: Gutzwiller; solid line: uniform approximation. 
            (c) Same as (b) but with actions scaled to be 10 times larger.
            }
            {Bifurcation}

In Fig.\ \ref{Bifurcation}(b), the contribution of these orbits to the
conductance is plotted. The dotted line gives the result of the trace
formula Eq.\ (\ref{traceFormula}). The amplitudes are diverging 
(arrows!) at the
bifurcations. The uniform approximation (solid line) removes the 
divergences. Fig.\ \ref{Bifurcation}(c) represents the corresponding 
data for a system scaled to have 10 times larger actions, thus being 
closer to the semiclassical limit. It is important now to note that 
the amplitudes in the uniform approximation are nearly constant over 
the bifurcations. We have thus shown that the bifurcations have no 
locally dominant influence on the conductance of the present system 
\cite{foot:hbar}.

\PostScript{2.5}{-0.4}{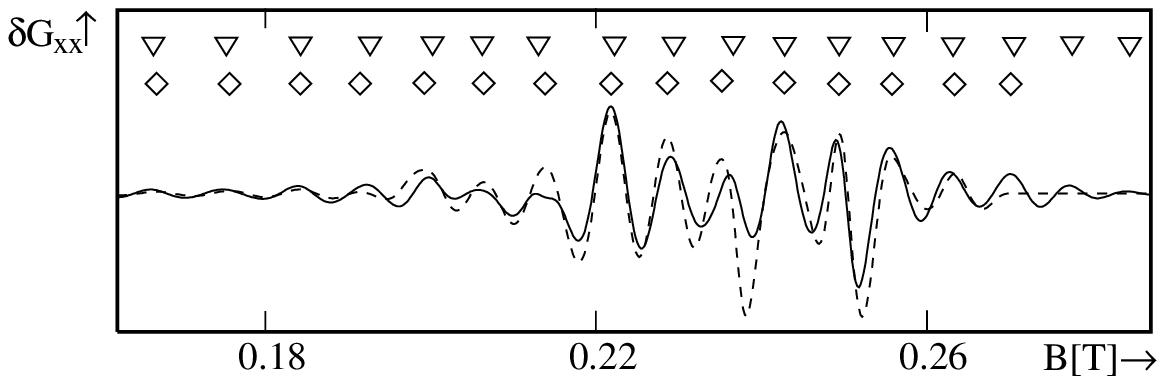}{-0.1}
            {$\delta G_{xx}$ using Eq.\ (\ref{traceFormula})
            (dashed line) and the uniform approximation (solid line)
            after a convolution over $B$. 
            Maxima are marked by boxes and
            triangles, respectively.
            }
            {Gxx}

Having established this result, we can further 
simplify our semiclassical treatment.
Whereas for individual orbits a uniform treatment of the bifurcations is 
vital, their influence becomes smaller if a larger number of orbits is
included. This is demonstrated in Fig.\ \ref{Gxx}, where the total
$\delta G_{xx}$ has been calculated for $s_d=1.86$ including all
relevant ($\sim$ 60) periodic orbits. The solid line shows the result 
of the uniform approximation, whereas the dotted line corresponds to the
standard Gutzwiller approach. To remove the spurious divergences 
in Eq.\ (\ref{traceFormula}) (and those due to bifurcations of higher 
codimension in the uniform approach), we have additionally convoluted 
$\delta G_{xx}$ over the magnetic field $B$ (cf.\ Ref.\ \cite{Bdisk}). 
The results are very similar \cite{Blaschke:PE97}. In particular, the
maximum positions 
are practically identical. In the following, we therefore use simply 
Eq.\ (\ref{traceFormula}) with an additional convolution over $B$.

The semiclassical result for the maximum positions in $\delta G_{xx}$ 
is shown in Fig.\ \ref{Results}(b). We do not obtain a detailed 
quantitative agreement with the experimental data, since no effort has
been made to optimize the model potential. Qualitatively, however, all
features of the observed phase plot in Fig.\ \ref{Results}(a) are  
reproduced. The spacing of the maxima will be analyzed in our extended
publication, where we also compare our results to those of 
quantum calculations, optimize the potential $V(r)$ and discuss the
scaling properties of our results. Presently we want to concentrate on
the dislocations (see the boxes in Fig.\ \ref{Results}), which are 
clearly reproduced in our approach. The semiclassical description even
reproduces quantitatively the local behavior at the dislocations. 
This is shown in Fig.\ \ref{Results}(c) that corresponds to the heavy
boxes in Figs.\ \ref{Results}(a) and (b). The points give the 
experimental maximum positions; 
the lines correspond to the semiclassical results (with slightly 
shifted but unscaled values of $s_d$ and $B$).
 
To understand the semiclassical origin of these dislocations, we consider 
for the moment a model system with only 7 closely related orbits. 
The inserts in Fig.\ \ref{Maxima89}(b) show Tr$\widetilde{\rm M}$ of 
these orbits versus $B$ for two different antidot diameters. With 
decreasing $s_d$, new orbits are born at the bifurcations. We classify 
the orbits in a grandparent, a parent and a child generation, depending 
if they are offsprings of orbit 1, 2, or 3, respectively. All members 
within a generation behave nearly identically, thus

\PostScript{6.3}{-0.6}{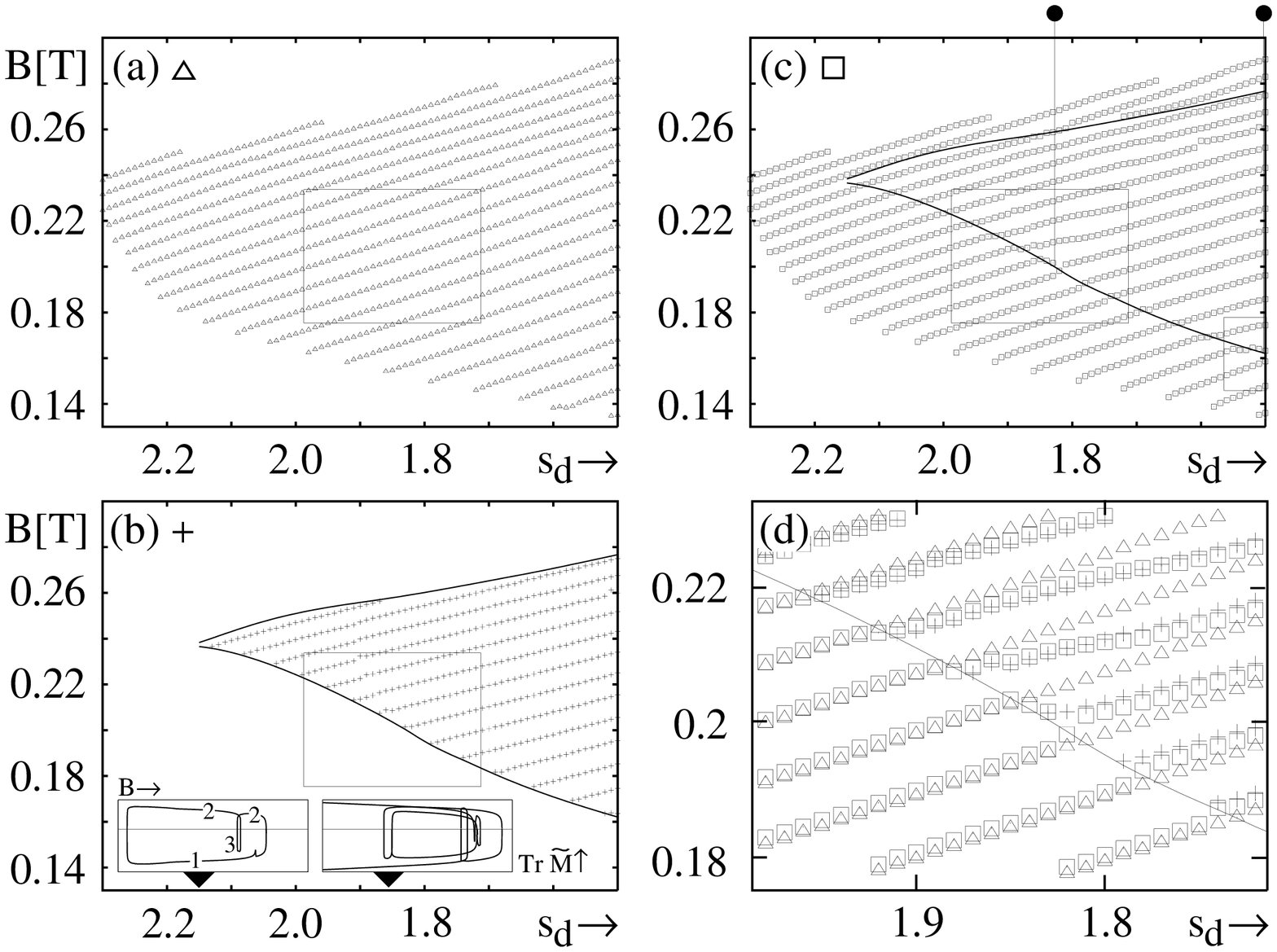}{-0.2}
            {Maximum positions due to different orbit 
             generations: (a) grandparents, (b) children, (c) all 
             generations. (d) Blow-up from (a)-(c): the maxima of the 
             total $\delta G_{xx}$ 
             (squares) follow the maxima of the children (crosses),
             where these exist, and those of the grandparents 
             (triangles) otherwise. Heavy lines indicate the loci
             of bifurcations in the ($s_d,B$) plane.
             }
            {Maxima89}

\noindent
justifying our 
classification. The contribution of the grandparent and the child 
orbits to the conductance is shown in Figs.\ \ref{Maxima89}(a) and (b), 
respectively. The behavior of each generation is in complete agreement
with the simple AB picture discussed above, but the effective areas 
and their dependence on $B$ are different. The children have a larger 
semiclassical amplitude than the grandparents, thus dominating the 
conductance. Therefore the maxima of the total $\delta G_{xx}$ closely
follow the children's wishes where they exist, and the grandparents' 
will otherwise, as becomes clear from Fig.\ \ref{Maxima89}(d). 
The parents' influence was found to be negligible 
throughout. The different orbit generations lead to slopes and 
spacings of the maxima that do not match along the generation 
boundaries. This is the origin of the observed dislocations which 
occur, indeed, close to the bifurcation lines. 

In the full calculation with over 60 orbits, the various families
with their bifurcation structures (gray lines in Fig.\ \ref{Results}b)
are superimposed. Only those dislocations survive for which the above 
model scenario is locally dominating and no further orbits interfere.
As a result, some of the dislocations disappear, some are slightly
shifted in the $(s_d,B)$ plane, and no unique 
one-to-one relation between dislocations and bifurcations can be 
established. Nevertheless, the qualitative pattern remains the same.

In summary, our semiclassical description successfully reproduces 
all main features observed experimentally in the magnetoconductance 
of a mesoscopic channel with antidots. We have analyzed especially 
the dislocations of the conductance maxima as functions of magnetic
field $B$ and antidot diameter $s_d$, and show that these are related 
to bifurcations of the leading classical periodic orbits of the system. 
The dislocations are due to the fact that the bifurcations define 
the border lines between regimes of different predominant orbit
generations, leading to different dependences of the conductance maxima 
on $B$ and $s_d$. This induces the observed dislocations of the maximum
positions, analogously to lattice defects at interfaces. As the 
classical dynamics are not affected by a rescaling of the system, 
the scaling behavior of the dislocations can be easily understood in 
the semiclassical approach.

The main mechanism for generating the dislocations has been demonstrated
for 7 model orbits forming three generations related through two
bifurcations, whereby the middle generation is least influential.
For individual orbits, a uniform semiclassical treatment of
the bifurcations is essential. For the total contribution of over
60 orbits, some cancellations take place and a convolution of the
trace formula (\ref{traceFormula}) over $B$ was found to be sufficient.

The ways in which the orbit bifurcations affect the quantum oscillations 
here is quite different from those reported in Refs.\  
\cite{RTD:all,Matsuy,Mag:bif}. There the relevant bifurcations
lead to period doublings, whereas here the
periods of all relevant orbits are approximately constant. Furthermore, 
in the RTD only a few orbits were found to be important, whereas the 
present system is dominated by a much larger number of orbits with
nearly identical actions, periods and amplitudes. It is not a
local enhancement of the amplitudes of isolated bifurcating orbits, 
but the occasional mismatch of the slowly varying contributions from competing 
orbit generations under the variation of the system parameters that 
causes the dislocations in the phase plots of the conductance maxima. 

We thank A. Sachraida, C. Gould and P. J. Kelly for providing us with 
the experimental data and helpful comments, and
S. Tomsovic for a critical discussion.


\begin{references}
\bibitem{foot:delta}The two most important cases are $\delta=1/6$ for
   tangent bifurcations and $\delta=1/4$ for period-doubling bifurcations.
\bibitem{foot:hbar} Leading-order orbits in 
    $\hbar$ need not dominate alone the quantum oscillations, as pointed 
    out for the equilateral triangular billiard \cite{Brack:Buch} 
    where an isolated orbit of relative order $\hbar^{1/2}$ is not 
    negligible. Similarly, for the circle billiard in a magnetic 
    field, skipping and cyclotron orbits of different 
    orders in $\hbar$ give comparable contributions \cite{Bdisk}.
\end{references}
\end{document}